\documentstyle[fleqn]{article}
\def\eq{equation}
\def\eqs{equations}

\begin{document}
\begin{center}
{\large\bf OHM'S LAW IN GENERAL RELATIVITY\\ AND
       CORIOLIS FORCE EFFECTS\\ IN ROTATING CONDUCTORS}\\
 \vspace*{0.5cm}

{\large\bf B.J. Ahmedov}\footnote{E-mail:
ahmedov@astrin.uzsci.net}\\
\vspace{0.5cm}

{\em
   Institute of Nuclear Physics and Ulugh Beg Astronomical
Institute\\
                  Ulughbek, Tashkent 702132, Uzbekistan}\\
\end{center}

\vspace{1cm}
\begin{abstract}

{It has been shown that a magnetic field proportional to angular
velocity of rotation $\omega$ arises around a rotating conductor with the
radial gradient of temperature $\nabla _r T$. However, the theoretical
value of the proportionality coefficient $10^{-17} {\rm cm}^{5/2}\cdot
{\rm g}^{1/2}\cdot {\rm deg}^{-1}$ does not coincide with the experimental
one $10^{-8} {\rm cm}^{5/2}\cdot {\rm g}^{1/2}\cdot {\rm deg}^{-1}$ [1].
At least two additional mechanisms act to produce an observable magnetic
field in the experiment [1].  First, the fact that the rotation of the
conductor is slowed down and second, the nonstationarity of the
temperature gradient during the measurements contribute to the vertical
magnetic field. The value of the magnetic flux arising from the azimuthal
current induced by the ``Coriolis force'' effect on the thermoelectric
radial current, is in good agreement with the experimental data [1].}

\end{abstract}

\newpage

Vasil'ev [1] has recently measured that a vertical
magnetic field is observable around
a rotating neutral conductor if one introduces a temperature gradient
perpendicular to the rotation axis. In his original analysis, the source of
this magnetic field is a steady-state temperature gradient.
However, we show here in the framework of electrodynamics of conducting
media in a rotating frame of reference that a permanent radial
temperature gradient produces
a very weak magnetic field, much less than that observed in the experiment.
On the contrary, since the heater was removed from the sample during the
measurements, a time-dependent radial temperature gradient and consequently
an electric current along the radius of conductor was set up.
From our analysis, an azimuthal current arising from the ``Coriolis force''
experienced by the radial current, was the source of the experimentally
detected vertical magnetic field.

In order to solve an electrodynamic problem for a conducting medium in
a noninertial frame of reference, in particular, in a rotating one, we
have to use the generally covariant Maxwell equations
\begin{eqnarray}
e^{\alpha\beta\mu\nu} F_{\beta\mu ,\nu}= 0,\hspace{10mm}
{H^{\alpha\beta}}_ {;\beta}= \frac{4\pi}{c}J^\alpha
\end{eqnarray}
with the constitutive relations between the field $F_{\alpha\beta}$
and the induction $H_{\alpha\beta}$ tensors of the electromagnetic field
\begin{eqnarray}
H_{\alpha\beta}=\frac{1}{\mu}F_{\alpha\beta} + \frac{1-\epsilon\mu}{\mu}
\big(u_\alpha F_{\sigma\beta}-u_\beta F_{\sigma\alpha}\big)u^\sigma,\\
F_{\alpha\beta}=\mu H_{\alpha\beta} + \frac{\epsilon\mu -1}{\epsilon}
\big(u_\alpha H_{\sigma\beta}-u_\beta H_{\sigma\alpha}\big)u^\sigma
\end{eqnarray}
and between the conduction current $\hat\jmath^\alpha$
and the field characteristics through
the thermoelectric power $\alpha$, the electric conductivity $\lambda$,
the electric permittivity $\epsilon$ and the magnetic permeability $\mu$.
Here the four-current $J^\alpha= c\rho_0 u^\alpha +\hat\jmath^\alpha,\quad
 \hat\jmath^\alpha
u_\alpha\equiv 0, u^\alpha$ is the four-velocity of the conductor,
$\rho _0$ is the proper density of electric charge.

The last relation (the general-relativistic Ohm's law)
can be obtained by modelling the equations of motion for conduction
electrons under the influence of both stationary gravitational and
electromagnetic
fields. The equation of motion for normal electrons can be written as
\begin{eqnarray}
mc^2u^\alpha _{(n);\sigma }u^\sigma _{(n)}=-eF^{\alpha\beta}u_{(n)\beta}+
\frac{ne^2}{\lambda}v^\alpha ,
\end{eqnarray}
where the last term results from the resistance force for conduction
electrons from the continuous medium,
$v^\alpha /c=\sqrt{1-v^2/c^2}u^\alpha -u^\alpha_{(n)}$ is the relative
velocity of positive ions and conduction electrons with the four velocity
$u^\alpha _{(n)}$, and $n$ is the density of conduction electrons.

After some transformations, according to [2], we can write the spatial part
of this equation as follows:
\begin{eqnarray}
mc\partial _Tv^\alpha =-eF^{\alpha\beta}u_\beta \nonumber\\
- e(F^{\alpha\sigma}+
F^{\rho\sigma}u_\rho u^\alpha )\frac{v_\sigma /c}{\sqrt{1-v^2/c^2}}+
\nonumber\\
+ \frac{ne^2}{\lambda}\frac{v^\alpha}{\sqrt{1-v^2/c^2}}-mc^2w^\alpha-
\frac{2mcv^\beta}{\sqrt{1-v^2/c^2}}A^\alpha _{.\beta}
\end{eqnarray}
where $\partial _Tv^\alpha$ denotes the time derivative of the relative
velocity [2],
$A_{\beta\alpha}= u_{[\alpha ,\beta ]}+u_{[\beta}w_{\alpha ]}$
is the relativistic rotation rate,
$w^\alpha$ is the absolute acceleration of the conductor, and
square brackets denote antisymmetrization.

Taking into account the fact that a stationary process, $\partial _T
v^\alpha=0$, establishes within the conductor in an extremely short time,
($\tau\approx 10^{-8}$ s), we obtain the equation of motion of electrons
which gives general-relativistic Ohm's law in its final form
\begin{eqnarray}
F_{\alpha\beta}u^\beta=\frac{1}{\lambda}\hat\jmath_\alpha+
R_H(F_{\nu\alpha}+u_\alpha u^\sigma F_{\nu\sigma})\hat\jmath^\nu
\nonumber\\ \quad
-aw_\alpha +
\stackrel{\perp}{\nabla}_\alpha \mu_e+\alpha\stackrel{\perp}{\nabla}_\alpha T
-b\hat\jmath^\beta A_{\alpha\beta}
\end{eqnarray}
if we add the terms due to
variations in both the chemical potential $\mu _e$ and the temperature $T$.
The last term in \eq (6) is responsible
for the Coriolis effect for the conduction current.
Here $R_H$ is the Hall constant,
$a=mc^2/e$  and $b=2mc/ne^2$, $\stackrel{\perp}{\nabla}_\alpha$
denotes the spatial part of a covariant derivative.

Ohm's law in the fields of gravity and inertia has been studied
theoretically in the recent papers [3-6], but the effect
of a ``Coriolis force'' upon an electric current was neglected.

Assume that a hollow cylindrical conductor (with $r_1$  and
$r_2$ as the radii of the inner and outer surfaces) with the radial
gradient of temperature $\nabla_rT$ is at rest in the rotating frame of
reference:
\begin{eqnarray}
     ds^2=-(c^2-\omega^2 r^2)dt^2+2\omega r^2d\varphi dt \nonumber\\
         + dr^2+r^2d\varphi^2+dz^2.
\end{eqnarray}
Let us further assume that the temperature gradient is constant as
proposed in [1]. Then the conduction current is absent inside the
conductor due to the stationarity of the situation and the finiteness of its
size. As a rigorous consequence of Ohm's law (6) we have the inner
electric field
\begin{equation}
E_r=A\frac{\omega^2 r}{c^2-\omega^2 r^2}+\alpha\frac{\partial T}
{\partial r},
\end{equation}
which exists inside the conductor, where
the parameter $A=a-\gamma M_ac^2/e$ [5],
$M_a$ is the atomic mass and $\gamma$ is a parameter of order $0.1$.
Solution of \eqs (1) gives the following expressions for the
space charge:
\begin{eqnarray}
\rho_0=\frac{1}{4\pi}\Biggl\{ \frac{\epsilon A\omega^2(2c^2+
\omega^2r^2)}{(c^2-\omega^2 r^2)^2}\nonumber\\  +\frac{\omega^2
r}{c^2-\omega^2 r^2} \frac{\partial(\epsilon A)}{\partial r}
\nonumber\\  + \frac{\epsilon\alpha} {r(1-\omega
^2r^2/c^2)}\frac{\partial T}{\partial r}+\frac{\partial (\epsilon
\alpha\partial T/\partial r)}{\partial r}\Biggr\}
\end{eqnarray}
and the surface charge distributions
$$
\sigma_{|r=r_1}=\frac{\epsilon A}{4\pi}\frac{\omega^2 r_1}{c^2-\omega^2r_1^2}
+\frac{\epsilon\alpha}{4\pi}\frac{\partial T}{\partial r}|_{r=r_1},
      \qquad \eqno (10a)
$$
 $$
\sigma_{|r=r_2}=-\frac{\epsilon A}{4\pi}\frac{\omega^2r_2}{c^2-\omega^2r_2^2}
-\frac{\epsilon\alpha}{4\pi}\frac{\partial T}{\partial r}|_{r=r_2}
      \qquad \eqno (10b)
$$
\setcounter{equation}{10}
at the inner and outer surfaces of the cylinder, respectively.

The magnetic field detected by an inertial observer is given by
\begin{equation}
B_z=-A\frac{\omega ^3r^2/c^3}{(1-\omega ^2r^2/c^2)^{3/2}}-\frac{\omega r}
{\sqrt{c^2-\omega ^2r^2}}\alpha\frac{\partial T}{\partial r}.
\end{equation}
Thus the magnetic flux through the inertial magnetometer is equal to
\begin{eqnarray}
\Phi =-\int_0^{2\pi }\int_0^{r_2}A\frac{\omega
^3r^3/c^3}{(1-\omega ^2r^2/c^2)^{3/2}}drd \varphi \nonumber\\
-\int_0^{2\pi }\int_0^{r_2}\frac{\omega r^2}{ \sqrt{c^2-\omega
^2r^2}}\alpha \nabla _rTdrd\varphi\nonumber\\   \approx
-\frac{A\pi }2\omega ^3r_2^4/c^3-\frac{2\pi \omega
r_2^3}{3c}\alpha \nabla _rT.
\end{eqnarray}
The part of the magnetic flux proportional to $\omega$ is given by
$\Phi =C(\nabla _rT)\omega .$ For typical values of the parameters
$r_2=3$ cm, $\alpha =10^{-7}\ {\rm cm}^{1/2}\cdot {\rm g}^{1/2}\cdot
s^{-1}\cdot {\rm deg}^{-1}$ the calculated value of the parameter $C$ lies
in the range $10^{-16} - 10^{-17}\ {\rm cm}^{5/2}\cdot {\rm g}^{1/2}\cdot
{\rm deg}^{-1}$ and does not coincide with the experimental one, $10^{-8}\
{\rm cm}^{5/2}\cdot {\rm g}^{1/2}\cdot {\rm deg} ^{-1}$ [1].

From our point of view, the above noncoincidence between the
theoretical and experimental results is due to the nonstationarity
of the processes in the analyzed experiment [1]. Two additional
mechanisms act to produce an observable magnetic field: (i) due to
the Stewart-Tolman effect, the rotation velocity decrease during
the measurements creates a magnetic field $2a \omega /c$ [7] and
the magnetic flux $\Phi =C_1\omega$, where $C_1\approx 10^{-6}\
{\rm cm}^{3/2}\cdot {\rm g}^{1/2}$; (ii) the nonstationary of the
gradient of temperature in the experiment [1] produces a radial
current $\hat\jmath _r=\lambda \alpha\nabla _rT$ (here we have
ignored the magnetic field arising from the radial current because
it does not depend on the angular velocity of rotation). According
to (6), there is a ``Coriolis force'' effect for the conduction
current and therefore an azimuthal current
\begin{equation}
\hat\jmath _\phi =b\lambda ^2\alpha(\nabla
_rT)\frac{\omega}{\sqrt{c^2- \omega ^2r^2}}
\end{equation}
arises and creates a vertical magnetic field
\[
H_z\approx 4\pi b\lambda ^2\alpha (\nabla _rT)\omega r/c^2.
\]
So the magnitude of the magnetic flux due to the ``Coriolis force''
is expected to be $\Phi =C_2(\nabla _rT)\omega$, where
$C_2\approx 10^{-8}\ {\rm cm}^{5/2}\cdot {\rm g}^{1/2}\cdot {\rm deg}^{-1}$
(if $\lambda = 5\cdot 10^{17}\ {\rm s}^{-1}$, $n\approx 10^{22}\
{\rm cm}^{-3}$), which is in good agreement with the experiment [1].

The magnetic field strength is determined by the sign of the thermoelectric
power $\alpha $ and therefore we can suggest that one might make
measurements for a cylinder
consisting either of layers of different materials (on height)
which have equal values (but opposite signs)
of thermoelectric powers or of material with
$\alpha =0$ (such as Mg, Pb). Since the total thermoelectric power
$\alpha$ for these samples
is zero, one can expect that the effect under study should vanish for them.

It should be concluded that several concurent effects have
appeared in the experiment. In order to separate them, some experiments
under different circumstances can be performed, viz. (i) under
constant rotation velocity and varying one of the gradients of
temperature; (ii) vice versa and (iii) when the velocity of rotation and
the gradient of temperature are constant.

\section*{Acknowledgments}

The author acknowledges financial support and hospitality at the
Abdus Salam International Centre for Theoretical Physics, Trieste,
and thanks A.V. Khugaev for helpful discussions. This research is
also supported in part by the UzFFR (project 01-06) and projects
F.2.1.09, F2.2.06 and A13-226 of the UzCST.

\end{document}